\documentclass{elsart41}
\usepackage{graphics}
\usepackage{graphicx}
\usepackage{amssymb}
\newcommand{\mib}[1]{\mbox{\boldmath$#1$}}

\begin{document}

%. 4 printed pages
\begin{frontmatter}

% Title, authors and addresses

% use the thanksref command within \title, \author or \address for footnotes;
% use the corauthref command within \author for corresponding author footnotes;
% use the  command for the email address,
% and the form \ead[url] for the home page:
% \title{Title\thanksref{label1}}
% \thanks[label1]{}
% \author{Name\corauthref{cor1}\thanksref{label2}}
% \ead{email address}
% \ead[url]{home page}
% \thanks[label2]{}
% \corauth[cor1]{}
% \address{Address\thanksref{label3}}
% \thanks[label3]{}

\title{Diffraction from Ordered States of Higher Multipoles 
}
%---- Don't remove this comment line! ----
%
% use optional labels to link authors explicitly to addresses:
% \author[label1,label2]{}
% \address[label1]{}
% \address[label2]{}

\author{Y. Kuramoto\corauthref{Kuramoto}},
\ead{kuramoto@cmpt.phys.tohoku.ac.jp}
\author{H. Kusunose},\,
\author{A. Kiss}

\address{
Department of Physics, Tohoku University, Sendai 980-8578,
Japan}

\corauth[Kuramoto]{Corresponding author. Tel: 81-22-795-6435, 
Fax: 81-22-795-6435}

\begin{abstract}

Possible ways of identification are discussed 
of an electronic order of higher multipoles such as octupoles and hexadecapoles. 
A particularly powerful method is resonant X-ray scattering (RXS)  using quadrupolar resonance processes called E2.
The characteristic azimuthal angle dependence of Ce$_{0.7}$La$_{0.3}$B$_6$ 
is interpreted as evidence of antiferro-octupole order. 
For PrRu$_4$P$_{12}$, eightfold pattern against azimuthal angle is predicted if its metal-insulator transition is a consequence of a hexadecapole order.   
In non-resonant superlattice Bragg scattering, hexadecapole contribution may also be identified because of absence of quadrupole component. 

\end{abstract}


\end{frontmatter}

\noindent
\textbf{\textit{I. Introduction}}\ \ \
In this paper, we shall discuss how to observe
electronic orders of higher multipoles such as octupoles and hexadecapoles ($2^4$-poles). 
These orders should of course result in anomaly of specific heat as in ordinary phase transition, but otherwise make it hard to identify the order parameter.  
Multipole orders have attracted growing interest recently.
In particular, 
Ce$_x$La$_{1-x}$B$_6$ exhibits a rich $H$-$T$ phase diagram with phases ranging from I to IV \cite{Hiroi97,Tayama97}.
The $\Gamma_{5u}$ octupole ordering model 
explains most of the characteristic behaviors in phase IV \cite{Kubo04}, such as the cusp-like behavior in the uniform susceptibility \cite{Tayama97}, absence of magnetic Bragg reflection in neutron scattering \cite{Iwasa03}, huge softening of $C_{44}$ mode in the strain susceptibility \cite{Suzuki98}, and minute lattice distortion along [111] \cite{Akatsu03}.

On the other hand, PrRu$_4$P$_{12}$ shows a metal-insulator transition with breathing type staggered lattice distortion \cite{sekine} .  The crystal symmetry remains cubic in the low temperature phase \cite{iwasa05}.  
The magnetic moment is unlikely to participate in the transition which is insensitive to magnetic field, and at which the susceptibility shows no distinct anomaly.    The large anomaly in the specific heat, on the other hand, requires involvement of $f$-electron degrees of freedom.
Thus it has been pointed out \cite{kuramoto05,takimoto} that a particular type of hexadecapole moment can be a candidate of the electronic order parameter in PrRu$_4$P$_{12}$.
In order to reproduce the CEF splittings below the transition, however, additional contribution of hexacontatetrapoles ($2^6$-poles) should be considered \cite{kiss05}. 

In the following, we shall demonstrate that octupole and hexadecapole orders 
show up as a characteristic diffraction property of X-ray scattering.
Especially powerful is identification of a multipole through the dependence on azimuthal angle.

\medskip\noindent
\textbf{\textit{
II. Multipoles under CEF}}\ \ \
We begin with properties of multipoles under crystalline electric field (CEF),
which is specified in terms of the one-body potential.  
Under the tetragonal point-group, which is relevant to filled skutterudites,
the CEF potential is written as
\begin{eqnarray}
V_{\rm CEF} 
&=& A_4 [O_4^0+5O_4^4]+A_6^{\rm c}[O_6^0-21O_6^4]+A_6^{\rm t}[O_6^2-O_6^6]
\nonumber \\
& \equiv & W\left[ x \frac{O_4}{60} +(1-|x|)\frac{O_6^c}{1260}+y\frac{O_6^t}{30}
 \right],
\label{V_CEF}
\end{eqnarray}
in the standard notation\cite{THY} where the superscript $c$ and $t$ mean cubic and tetragonal, respectively.
In the case of $f$ electrons, their orbital angular momentum $l=3$ gives the upper limit of the rank $L=2\times 3= 6$ for the CEF potential.  
In the cubic point group $O_h$, the term $yO_6^{t}$ does not appear.
The operator $O_L^M$
is proportional to the sum of spherical tensors 
$T_M^{(L)}$ and $T_{-M}^{(L)}$.
The particular combinations of $O_L^M$ with the coefficients $A_L$ transform as scalars in the point group.  These scalars appear  after the intensity parameter $W$.  

It is obvious from this form that a scalar in the point group is in general a multipole in the spherical symmetry.   
The anisotropy associated with a multipole is probed by such experiments that are sensitive to detailed shape of wave functions.
In microscopic terms, the shape specifies distribution of charge and spin of electrons.
The magnetic octupole moment, for example, consists of distribution of magnetic dipole moments that add up to zero around the rare-earth site.   Both spin and orbital angular momentum can be the origin of microscopic magnetic moment. 
The electric quadrupole moment, on the other hand, can be nonzero for non-spherical distribution of charges.  
The hexadecapole becomes the lowest multipole moment for certain non-spherical distribution of charges.

In terms of the coordinates $\hat{\mib r}=(x, y, z)$ on the unit sphere 
for describing angle variables, a set of octupoles
under the cubic symmetry is given by  
\begin{equation}
\Gamma_{4u}: \ x(5x^2-3), \ 
y(5y^2-3), \
z(5z^2-3), \
\end{equation}
which transforms in the same way as the set $x,y,z$.  
Thus the dipole and a part of octupole operators are mixed.  
The irreducible representation is labeled $\Gamma_{4u}$.
On the other hand, another set of octupoles
\begin{equation}
\Gamma_{5u}: \ x(y^2-z^2), \ 
y(z^2-x^2), \ 
z(y^2-z^2), \ 
\end{equation}
does not mix with dipoles, and are labeled $\Gamma_{5u}$.
Among even rank tensors,
quadrupole and a part of hexadecapole $(L=4)$ operators mix under the cubic symmetry.  Namely, a set of hexadecapoles
\begin{equation}
\Gamma_{5g}: \ xy(7z^2-1), \ 
yz(7x^2-1), \ 
zx(7y^2-1), \ 
\end{equation}
transforms in the same way as $xy, yz, zx$ and belongs to the $\Gamma_{5g}$ representation.

Some cases have been found where the CEF states allow these multipoles to fluctuate because of their degeneracy, or because of small enough CEF splitting.  A typical example is the four-fold degenerate CEF ground state in CeB$_6$.  There is an orbital degeneracy in addition to the Kramers degeneracy in this case.  Then one can expect both magnetic and orbital orderings in the ground state, unless quantum fluctuation like the Kondo effect blocks the order.  Actually a magnetic order called phase III is realized in the ground state of pure CeB$_6$, but the Kondo effect becomes dominant as the concentration of Ce decreases in Ce$_{1-x}$La$_x$B$_6$.

\medskip\noindent\textbf{\textit{
III. Non-resonant X-ray scattering from multipoles}}\ \ \
The most straightforward method
to probe a multipole order is non-resonant X-ray scattering \cite{Lovesey05}.
The scattering amplitude is proportional to the form factor
\begin{equation}
F(\mib{q}) =
\langle 
\sum_j \exp(i\mib{q}\cdot\mib{r}_j) \rangle ,
\end{equation}
where $\mib r_j$ is the coordinate of $j$-th electron in a unit cell.
The structure factor is nonzero only for Bragg vectors
or a superlattice vector $\mib q=\mib Q$ involving the charge density.
For a given $\mib{q}$, the plane wave is expanded in terms of spherical harmonics times spherical Bessel functions.   
Taking the $z$-axis of spherical harmonics along direction of $\mib q$,  
we obtain 
\begin{equation}
F(\mib{q}) = \sum_{n=0}^3 c_n 
\langle j_{2n}(q)\rangle 
\langle T_{0}^{(2n)}\rangle_{\mib{q}}, 
\label{nonresonant}
\end{equation}
where $c_{n}$ is a numerical factor, and 
$\langle T_{M}^{(L)}\rangle_{\mib{q}}$ with $M=0$ and $L\leq 6$
indicates that $\mib q$ is taken along the $z$-axis.
In the presence of an antiferro-orbital order with $\Gamma_{5g}$ symmetry, both $T_M^{(2)}$ and $T_M^{(4)}$ contribute to the scattering.  
Furthermore we have introduced the notation 
$\langle j_L(q)\rangle = \int_0^\infty dr  j_L(qr) f (r)r^2,
$
where $f(r)$ is the radial wave function of an $f$ electron, and 
$j_L(qr)$ is a spherical Bessel function of order $L$.
The scattering amplitude depends on direction of $\mib q$ relative to the cubic crystal axis through the spherical tensors.  On the other hand, the dependence on $q=|\mib q|$ comes from $\langle j_L(q)\rangle$  
Thus the scattering intensity gives information about relative weight of quadrupole and hexadecapole operators.  
The case of CeB$_6$ has been analyzed in ref. \cite{tanaka04a} which demonstrates that  the dominant weight comes in fact from the {\it hexadecapole} component.

\medskip\noindent\textbf{\textit{
IV. Resonant X-ray scattering from multipoles}}\ \ \
We now discuss resonant X-ray scattering (RXS).
Recent measurement has obtained information of the order parameter symmetry in phase IV 
in Ce$_{0.7}$La$_{0.3}$B$_6$ below $T_{\rm IV}=1.5$ K \cite{Mannix05}.
The superlattice Bragg reflections at 
${\mib Q}=(3/2,3/2,3/2)$ is taken 
with the rotation axis normal to the [111] surface.
Both non-rotated ($\sigma$-$\sigma'$), and rotated ($\sigma$-$\pi'$) polarization channels have been measured at the electric quadrupole (E2) resonance near the Ce $L_2$ absorption edge.
In the $\sigma$-$\sigma'$ channel, the scattering intensity exhibits six-fold oscillation, which indicates the occurrence of the electronic order with three-fold symmetry along [111].  In contrast, the $\sigma$-$\pi'$ channel shows three-fold oscillation.

It is demonstrated in ref. \cite{kusunose05} that the $\Gamma_{5u}$-type octupole order can reproduce the azimuthal angle dependences both in the $\sigma$-$\sigma'$ and $\sigma$-$\pi'$ channels.
We now outline the argument of ref. \cite{kusunose05}.
The RXS amplitude per unit cell
is given by the formula:  \cite{Lovesey05} 
\begin{equation}
F_{\rm reso}=-\frac{\Delta^2}{\hbar^2c^2}\sum_{m}\frac{W_{fi}^{(m)}}{\hbar\omega-\Delta+i\Gamma/2},
\label{RXS}
\end{equation}
where simplifying approximation has been made
that all intermediate states $(m)$ have the same energy and width $\Gamma$.
The energy of the incident photon $\hbar\omega$ is tuned close to the absorption edge $\Delta$ of the relevant atom.
The amplitude is proportional to 
\begin{eqnarray}
W_{fi}^{(m)}&=&\langle f| {\mib \epsilon}'\cdot{\mib P}|m\rangle\langle m|{\mib \epsilon}\cdot{\mib P}|i\rangle \nonumber \\
&& +\langle f| {\rm Tr}(\hat{X}'\cdot\hat{Q})|m\rangle\langle m|{\rm Tr}(\hat{X}\cdot\hat{Q}|i\rangle,
\end{eqnarray}
where the first term in the right-hand side comes 
from the electric dipole (E1) transition, and the second term from the quadrupole (E2) transition.
Here we have defined the dipole and the quadrupole operators for each ion as
$P^\alpha=e\sum_{n}^{\rm atom}r_n^\alpha$ and $\hat{Q}_{\alpha\beta}=e\sum_n^{\rm atom}r_n^\alpha r_n^\beta/2$, and the matrix made from the photon wave and polarization vectors, $\hat{X}_{\alpha\beta}=k^\alpha \epsilon^\beta/2$.
The prime represents quantities concerning the scattered radiation.
The maximum rank probed by the E2 transition is $L=4$.  Namely, the triakontadipole ($L=5$) and hexacontatetrapole cannot be detected by the RXS.

Table \ref{E2} summarizes the azimuthal angle dependences of multipole scattering in the E2 scattering
for the case where the crystal coordinate ${\mib R}=(X,Y,Z)$
is the same as the scattering coordinate ${\mib r}$.
The latter is defined by taking the scattering plane as the $yz$-plane, and the origin of $\psi$ along the $y$ axis  \cite{kusunose05}.
Here we have used the real  tensors for $M>0$ defined by
\begin{eqnarray}
T^{(L)}_{cM}&=&\frac{(-1)^M}{\sqrt{2}}\left[ T^{(L)}_M + T^{(L)*}_M \right],
\\
T^{(L)}_{sM}&=&\frac{(-1)^M}{\sqrt{2}i}\left[ T^{(L)}_M - T^{(L)*}_M \right].
\end{eqnarray}
The tensor $T^{(L)}_{cM}$ is proportional to the multipole operator $O_L^M$.
If the rotation axis for the azimuthal scan is different from the crystal axis $Z$, one has to relate the coordinates $\mib r$ and $\mib R$ by linear transformation.  For example, Table \ref{111} shows the result 
for the case where rotation axis of the azimuthal scan is $[111]$.  
If degeneracy remains in the irreducible representation $\Gamma$, we take a linear combination so that the  principal axis of the multipole is parallel to [111].

\begin{table}[t]
\caption{Angle dependence of RXS amplitude of octupoles $T^{(3)}$ and hexadecapoles $T^{(4)}$ for the E2 transition.
The elevation angle is written as $\xi = \pi/2-\theta$ with $\theta$ being the incident angle.}
\begin{tabular}{ccc}
\\ \hline\hline
 & $\sigma\sigma'$ & $\sigma\pi'$ \\ \hline
$T^{(3)}_{c3}$ & $\frac{1}{4}\sin 2\xi\cos 3\psi $ & $-\frac{1}{16}(\cos\xi+3\cos 3\xi )\sin 3\psi$ \\
$T^{(3)}_{c2}$ & 0 & $-\frac{1}{2}\sqrt{\frac{3}{2}}\cos^2\xi\sin\xi\cos 2\psi$ \\
$T^{(3)}_{c1}$ & $-\frac{1}{4}\sqrt{\frac{3}{5}}\sin 2\xi \cos\psi$ & $\frac{1}{8}\sqrt{\frac{3}{5}}\cos\xi(\cos 2\xi -3)\sin\psi$ \\
$T^{(3)}_{s3}$ & $-\frac{1}{4}\sin 2\xi \sin 3\psi$ & $-\frac{1}{16}(\cos\xi+3\cos 3\xi )\cos 3\psi$ \\
$T^{(3)}_{s2}$ & 0 & $\frac{1}{2}\sqrt{\frac{3}{2}}\cos^2\xi\sin\xi\sin 2\psi$ \\
$T^{(3)}_{s1}$ & $\frac{1}{4}\sqrt{\frac{3}{5}}\sin 2\xi \sin\psi$ & $\frac{1}{8}\sqrt{\frac{3}{5}}\cos\xi(\cos 2\xi -3)\cos\psi$ \\
$T^{(3)}_0$ & 0  & $\frac{1}{4\sqrt{10}}\sin\xi(3\cos 2\xi -1)$ \\ \hline
$T^{(4)}_{c4}$ & $-\frac{1}{2\sqrt{2}}\cos^2\xi\cos 4\psi$ & $-\frac{1}{2\sqrt{2}}\sin\xi\cos^2\xi\sin 4\psi$ \\
$T^{(4)}_{c3}$ & 0 & $\frac{1}{4}\cos^3\xi\cos 3\psi$ \\
$T^{(4)}_{c2}$ & $-\frac{1}{\sqrt{14}}\sin^2\xi\cos 2\psi$ & $\frac{1}{4\sqrt{14}}\sin\xi(\cos 2\xi -3)\sin 2\psi$ \\
$T^{(4)}_{c1}$ & 0 & $-\frac{1}{8\sqrt{7}}\cos\xi(3\cos 2\xi -5)\cos\psi$ \\
$T^{(4)}_{s4}$ & $\frac{1}{2\sqrt{2}}\cos^2\xi\sin 4\psi$ & $-\frac{1}{2\sqrt{2}}\sin\xi\cos^2\xi\cos 4\psi$ \\
$T^{(4)}_{s3}$ & 0 & $-\frac{1}{4}\cos^3\xi\sin 3\psi$ \\
$T^{(4)}_{s2}$ & $\frac{1}{\sqrt{14}}\sin^2\xi\sin 2\psi$ & $\frac{1}{4\sqrt{14}}\sin\xi(\cos 2\xi -3)\cos 2\psi$ \\
$T^{(4)}_{s1}$ & 0 & $\frac{1}{8\sqrt{7}}\cos\xi(3\cos 2\xi -5)\sin\psi$ \\
$T^{(4)}_0$ & $\frac{1}{4\sqrt{70}}(5-3\cos 2\xi )$ & 0 \\ \hline\hline
\label{E2}
\end{tabular}
\end{table}

\begin{table}%[b]
\caption
{Angle dependence of RXS intensity with [111] as the rotation axis.  The order parameter is rank $p$ multipoles with irreducible representation $\Gamma$
in the $O_h$ group.
The origin $\psi=0$ is taken in the $[111]$-$[11\bar{2}]$ plane.}
\begin{tabular}{ccc}
\\ \hline\hline
$p$-$\Gamma$ &  E2 $\sigma\sigma'$ & E2 $\sigma\pi'$ \\ \hline
3-2u & $\frac{1}{36}\sin^2 2\xi \sin^2 3\psi$ & $\frac{1}{144}(3\cos 2\xi -1)^2\cos^2\xi$ \\
 & & $\mbox{\hspace{0.3cm}}\times\left[\frac{1}{\sqrt{2}}\tan\xi-\cos 3\psi\right]^2$ \\
3-4u & $\frac{1}{36}\sin^2 2\xi \sin^2 3\psi$ & $\frac{1}{144}(3\cos 2\xi -1)^2\cos^2\xi$ \\
& & $\mbox{\hspace{0.3cm}}\times\left[\frac{1}{\sqrt{2}}\tan\xi+\cos 3\psi\right]^2$ \\
3-5u & $\frac{1}{16}\sin^2 2\xi \cos^2 3\psi$ & $\frac{1}{256}(\cos\xi+3\cos 3\xi )^2\sin^2 3\psi$ \\
4-4g & 0 & $\frac{1}{16}\cos^6\xi\cos^2 3\psi$ \\
4-5g & $\frac{1}{1512}(5-3\cos 2\xi )^2$ & $\frac{1}{1296}\cos^6\xi\sin^2 3\psi$ \\
\hline\hline
\label{111}
\end{tabular}
\end{table}

We define the amplitude $f_{5u}^{[111]}(\alpha)$ from the [111] domain for the 
scattering channel $\alpha \ (=\sigma, \pi)$.  The other three equivalent domains 
$\mu = [\bar{1}11], [1\bar{1}1], [11\bar{1}]$ have different amplitudes $f_{5u}^{\mu}(\alpha)$  in general.  The total intensity is proportional to 
\begin{equation} 
I_\alpha=A_\alpha\left[
w|f_{5u}^{[111]}(\alpha)|^2+\frac{1-w}{3}\sum_{\mu} 
|f_{5u}^{\mu }(\alpha)|^2\right]. 
\end{equation}
We take the angle $\xi=39^\circ$ as required from the incident photon energy and the scattering wave vector ${\mib q}$ \cite{Mannix05}.
We assume that the four equivalent domains are equally populated, i.e., $w=1/4$.
The three-fold oscillation is explained in terms of domains whose principal axis 
is other than [111].  
Each domain gives the scattering intensity dependent on azimuthal angle.
Since each domain is rotated by $2\pi/3$ around [111],
the sum of the intensity aquires the three-fold symmetry independent of details of contribution of each domain. 
The intensity factors are chosen as
$A_{\sigma\sigma'}=120$, $A_{\sigma\pi'}=70$. 
The difference between 
$A_{\sigma\sigma'}$ and $A_{\sigma\pi'}$ 
may be ascribed to 
extrinsic background from non-resonant contribution and/or different energies of intermediate states as noted below eq.(\ref{RXS}).
The fitting with these intensity factors reproduces semi-quantitatively the observed data.
Among all multipoles accessible by the RXS, only the $\Gamma_{5u}$-type octupole yields the maximum at $\psi=0$ in the $\sigma$-$\sigma'$ channel, which is consistent with the observed oscillation  \cite{Mannix05}.
The [111] domain predominates over the other domains in the $\sigma$-$\sigma'$ channel.
In the $\sigma$-$\pi'$ channel, on the other hand, 
the contribution from the [111] domain is much less than that from 
the other three equivalent domains.
As a  result, a threefold component appears with the maximum at $\psi=0$.
Therefore, the $\Gamma_{5u}$-type octupole in three-fold axis with the four equivalent domains explains the RXS result  in Ce$_{0.7}$La$_{0.3}$B$_6$.

\begin{figure}
\begin{center}
\includegraphics[width=8cm]{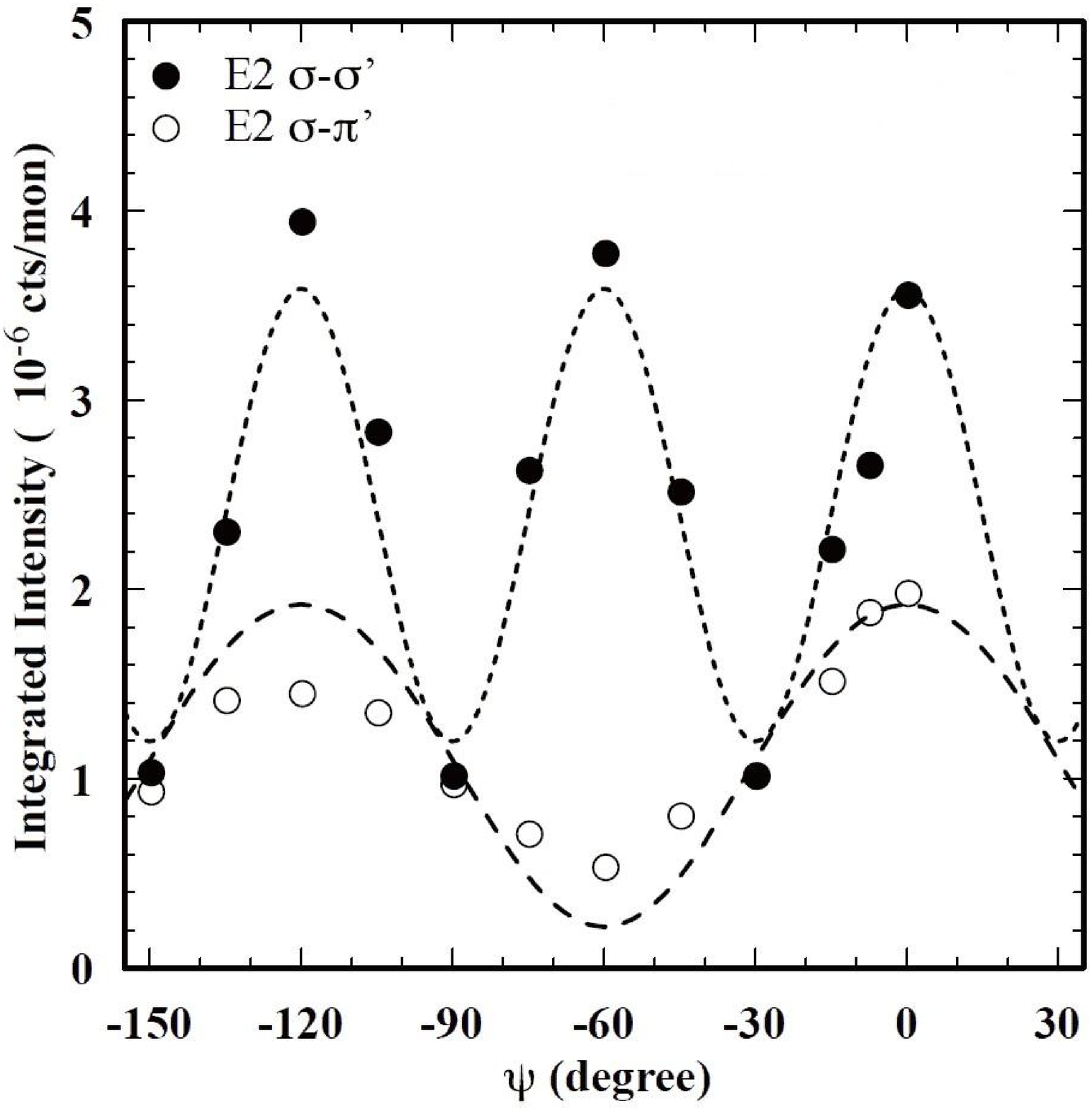}
\caption{Comparison of azimuthal angle dependence of RXS between theory and experiment.  The symbols are taken from Mannix {\it et al.} \cite{Mannix05}. The dotted and the dashed lines are the result for the $\Gamma_{5u}$-type octupole order as described in the text.}
\end{center}
\end{figure}

\medskip\noindent\textbf{\textit{
V. X-ray scattering from PrRu$_4$P$_{12}$
}}\ \ \
We consider how to check possible hexadecapole order transition in 
 PrRu$_4$P$_{12}$.
According to Table \ref{E2}, hexadecapoles can be probed by E2 scattering.  The combination $O_4^0+5O_4^4$ constitutes a scalar in the cubic symmetry.  
Here $O_4^4$ corresponds to $T_{c4}^4$ in Table \ref{E2}, and give rise to a characteristic dependence on azimuthal angle.  
The dependence $\cos 4\psi$ in the $\sigma$-$\sigma'$ and $\sin 4\psi$ in the $\sigma$-$\pi'$ channels are not shared by amplitude for other multipoles.  
In particular, there is no contribution from $O_4^0$ in the $\sigma$-$\pi'$ channel.
The resulting intensity has the eight-fold pattern ($\propto 1-\cos 8\psi$) with a minimum at  $\psi =0$.   The $\sigma$-$\sigma'$ channel will show a four-fold pattern by combination with $O_4^0$. 
In this way resonant X-ray scattering using the E2 transition can identify the hexadecapole order.    
On the other hand, the non-resonant scattering also has contribution from hexadecapoles as shown in eq.(\ref{nonresonant}).  
Although the monopole contribution with $c_0$ should dominate the intensity, 
one may hope to identify the contribution from $L=4$  since the quadrupole component $L=2$ should be absent.  
Identification of the $L=4$ component provides  alternative detection of the hexadecapole order.

In summary, we have discussed that higher multipoles can be detected by RXS in the quadrupole (E2) transition.  The observed azimuthal angle dependence in 
Ce$_{0.7}$La$_{0.3}$B$_6$ is consistent with the pure octupole order with the $\Gamma_{5u}$ symmetry, provided four equivalent domains are present in the sample.  If our interpretation is correct, the three-fold symmetry should not appear for a sample with a single domain.  
A new experiment is proposed to probe possible antiferro-hexadecapole order in PrRu$_4$P$_{12}$.

\medskip
This work was supported partly by 
Grants-in-Aid for Scientific Research on Priority Area ``Skutterudite" , and 
for Scientific Research (B)15340105 
of the Ministry of Education, Culture, Sports, Science and Technology, Japan.


\begin{thebibliography}{99} %% The number "99" means that this list has more than nine items.

\bibitem{Hiroi97} M. Hiroi, M. Sera, N. Kobayashi and S. Kunii: Phys. Rev. B {\bf 55} (1997) 8339.
\bibitem{Tayama97} T. Tayama, T. Sakakibara, K. Tenya, H. Amitsuka and S. Kunii: J. Phys. Soc. Jpn. {\bf 66} (1997) 2268.
\bibitem{Kubo04} K. Kubo and Y. Kuramoto: J. Phys. Soc. Jpn. {\bf 73} (2004) 216.
\bibitem{Iwasa03} K. Iwasa et al.
%, K. Kuwahara, M. Kohgi, P. Fischer, A. D\"omni, L. Keller, T. C. Hansen, S. Kunii, N. Metoki, Y. Koike and K. Ohoyama; 
Physica B {\bf 329}-{\bf 333} (2003) 582.
\bibitem{Suzuki98} O. Suzuki, T. Goto, S. Nakamura, T. Matsumura and S. Kunii: J. Phys. Soc. Jpn. {\bf 67} (1998) 4243.
\bibitem{Akatsu03} M. Akatsu, T. Goto, Y. Nemoto, O. Suzuki, S. Nakamura and S. Kunii: J. Phys. Soc. Jpn. {\bf 72} (2003) 205.
\bibitem{sekine} C. Sekine, T. Uchiumi, I. Shirotani, and T. Yagi: Phys. Rev. Lett. \textbf{79} (1997) 3218.
\bibitem{iwasa05} K. Iwasa et al. Phys. Rev. B{\bf 72} (2005) 024414.
\bibitem{kuramoto05}  Y. Kuramoto, J. Otsuki, A. Kiss and H. Kusunose, Proc. YKIS2004 (to be published in Prog.Theor.Phys.).
\bibitem{takimoto} T. Takimoto: preprint submitted to J.Phys.Soc.Jpn.
\bibitem{kiss05} A. Kiss and Y. Kuramoto: to be published.
\bibitem{THY} K. Takegahara, H. Harima, and A. Yanase: J. Phys. Soc. Jpn. \textbf{70} (2001) 1190.
\bibitem{Lovesey05} S.W. Lovesey et al: 
Physics Reports {\bf 411} (2005) 233.
\bibitem{tanaka04a} 
Y. Tanaka et al., 
Europhys. Lett. {\bf 68} (2004) 671. 
\bibitem{Mannix05} D. Mannix, Y. Tanaka, D. Carbone, N. Bernhoeft and S. Kunii: Phys. Rev. Lett. {\bf 95} (2005) 117206.
\bibitem{kusunose05} H. Kusunose and Y. Kuramoto: J. Phys. Soc. Jpn. {\bf 74} (2005) 3139.

\end{thebibliography}
\end{document}